\documentclass[12pt, fleqn]{article}
\usepackage{latexsym,amsfonts,amssymb}

\usepackage{amsfonts}
\usepackage{amssymb}
\usepackage{amsbsy}
\usepackage{amsmath}
\usepackage{epsf}

\newtheorem{theo}{Theorem}
\newcommand{\bt}{\begin{theo}}
\newcommand{\et}{\end{theo}}

\newcommand{\BEQ}{\begin{equation}}     
\newcommand{\BEA}{\begin{eqnarray}}
\newcommand{\BD}{\begin{displaymath}}
\newcommand{\be}{\begin{equation}}
\newcommand{\bel}[1]{\begin{equation}\label{#1}}
\newcommand{\EEQ}{\end{equation}}
\newcommand{\EEA}{\end{eqnarray}}
\newcommand{\ED}{\end{displaymath}}
\newcommand{\ee}{\end{equation}}

\newcommand{\demi}{\frac{1}{2}}         
\newcommand{\wit}[1]{\widetilde{#1}}    
\renewcommand{\vec}[1]{{\boldsymbol{#1}}} 

\def\bbbz{{\mathchoice {\hbox{$\sf\textstyle Z\kern-0.4em Z$}}
{\hbox{$\sf\textstyle Z\kern-0.4em Z$}}
{\hbox{$\sf\scriptstyle Z\kern-0.3em Z$}}
{\hbox{$\sf\scriptscriptstyle Z\kern-0.2em Z$}}}}

\catcode`\@=11
\def\numberbysection{\@addtoreset{equation}{section}
  \def\theequation{\thesection.\arabic{equation}}}
\numberbysection

\begin{document}
\begin{titlepage}

\medskip

\begin{center}
{\Large \bf The exotic conformal Galilei algebra and \\
nonlinear partial differential equations }
\end{center}

\centerline{{\bf Roman Cherniha}$^{a,b}$\footnote{\small e-mail:
{\tt cherniha@imath.kiev.ua}} and {\bf Malte
Henkel}$^b$\footnote{\small e-mail: {\tt henkel@lpm.u-nancy.fr} }}
\vskip 0.5 cm
\centerline{$^a$Institute of Mathematics, National Academy of Science of Ukraine}
\centerline{3, Tereshchenkivs'ka Str., UA - 01601 Kyiv, Ukraine}
\centerline{$^b$Groupe de Physique Statistique,}
\centerline{D\'epartement de Physique de la Mati\`ere et des Mat\'eriaux,
Institut Jean Lamour\footnote{Laboratoire associ\'e au CNRS UMR 7198},}
\centerline{CNRS -- Nancy Universit\'e -- UPVM, B.P. 70239,}
\centerline{F -- 54506 Vand{\oe}uvre l\`es Nancy Cedex, France}

\begin{abstract}
The conformal Galilei algebra ({\sc cga}) and the exotic conformal
Galilei algebra ({\sc ecga}) are applied to construct  partial
differential equations (PDEs) and systems of PDEs, which admit these
algebras. We show that there are no single second-order PDEs
invariant under the {\sc cga} but systems of PDEs can admit this
algebra. Moreover, a wide class of nonlinear PDEs exists, which are
conditionally invariant under {\sc cga}. It is further shown that
there are systems of non-linear PDEs admitting {\sc ecga} with the
realisation obtained very recently in [D. Martelli and Y. Tachikawa,
{\tt arXiv:0903.5184v2 [hep-th] (2009)]}. Moreover,  wide classes of
non-linear systems, invariant under two different  10-dimensional
subalgebras of {\sc ecga} are explicitly constructed and an example
with possible physical interpretation is presented.
\end{abstract}

\textbf{Keywords:} partial differential equation (PDE), the
conformal Galilei algebra, Lie symmetry,\\ conditional symmetry.

\vfill
J. Math. Anal. Appl. {\tt at press} (2010) 
\end{titlepage}

\section{Introduction}

Symmetries have since a long time played an important r\^ole in the
analysis of physical systems. In this paper, we consider
non-relativistic space-time symmetries. The best known of those is
the Lie algebra in $N$ spatial dimensions with the basic operators
\BEA {}
 X_{\pm 1,0}, \, Y_{\pm 1/2}^{(j)}, \,  M_0, \,
R_0^{(jk)} \;\; , \;\;
j,k=1,\ldots,N \label{0-1}
\EEA
The operators (\ref{0-1}) satisfy  the non-vanishing
commutation relations
\BEA
{} \bigl[ X_n, X_{n'}\bigr] &=& (n-n') X_{n+n'} \;\; , \;\; \hspace{0.33truecm}
{} \bigl[ X_n, Y_m^{(j)}\bigr] \:=\: \left(\frac{n}{2} -m\right) Y_{n+m}^{(j)}
\nonumber \\
{} \bigl[ Y_{1/2}^{(j)}, Y_{-1/2}^{(k)} \bigr] &=& \delta^{j,k}\,
M_0 \hspace{1.35truecm} \;\; , \;\; {} \bigl[ R_0^{(jk)},
Y_m^{(\ell)} \bigr] \:=\: \delta^{j,\ell}\, Y_m^{(k)} -
\delta^{k,\ell}\, Y_m^{(j)} \label{gl:schcr} \EEA where
$R_0^{(jk)}\in\mathfrak{so}(N)$, $j,k,\ell\in\{1,\ldots,N\}$,
$n,n'\in\{\pm 1,0\}$ and $m=\pm\demi$. Since  Sophus Lie \cite{lie}
discovered this algebra for $N=1$ as the maximal  algebra of
invariance (MAI) of the one-dimensional linear heat equation, this
algebra has been intensively studied and today is often called {\it
Schr\"odinger algebra}, to be denoted in this paper by
$\mathfrak{sch}(N)$. Besides being the Lie algebra of invariance of
the linear  heat (diffusion) and free Schr\"odinger equations, see
e.g. \cite{ovs,ol,Fush93} and references therein, it is also the Lie
symmetry algebra of non-linear systems of evolution equations and
Schr\"odinger type equations \cite{f-ch1,f-ch2,rideau-win}. Probably
the first  example
of a Schr\"odinger-invariant system of non-linear equations is given by  the
hydrodynamic equations of motion of a compressible fluid \cite{ovs}
\BEQ
\partial_t \rho + \vec{\nabla}\cdot\bigl( \rho \vec{v}\bigr) =0 \;\; , \;\;
\rho \bigl( \partial_t + (\vec{v}\cdot\vec{\nabla})\bigr)\vec{v} +
\vec{\nabla} P =0 \EEQ where
$\vec{\nabla}=(\partial_{r_{1}},...,\partial_{r_{N}})$ and $ \cdot$
means the scalar product, while  $\rho=\rho(t,\vec{r})$ is the
density, $\vec{v}=\vec{v}(t,\vec{r})$ is the velocity of the fluid.
Furthermore, the pressure $P$ satisfies a polytropic equation of
state $P= \rho^{\gamma}$ with a polytropic exponent $\gamma=1+2/N$.
Physicists have rediscovered this result not so long ago
\cite{Hass00}.

It is well-known that the Schr\"odinger algebra $\mathfrak{sch}(N)$
can be embedded into the (complexified) conformal Lie algebra
$\mathfrak{conf}(N+2)$ in $N+2$ dimensions \cite{Burdet73}. When one
considers explicit space-time representations which contain a
dimensional constant $c$ with the physical
units of velocity, taking the formal non-relativistic limit
$c\to\infty$ does not lead back to the Schr\"odinger algebra
$\mathfrak{sch}(N)$ but rather to a non-isomorphic Lie algebra,
which actually is a parabolic subalgebra of $\mathfrak{conf}(N+2)$ \cite{Henkel03}. This algebra
was identified at least as early as 1978
in \cite{hp78} and can indeed be obtained from the conformal algebra
$\mathfrak{conf}(N+1)$ by a group contraction. In the physical
literature, this algebra is usually called the {\it conformal
Galilei algebra} {\sc cga}$(N)$ \cite{hp78,negro-97}, but the name
of {\it altern algebra} \cite{Henk97} is also used, with reference
to the physical contexts (space-time geometry and ageing phenomena,
respectively), where it was identified, and it will be the object of
study in this paper.
The algebra {\sc cga}$(N)$ is spanned by the generators
\BEA
X_{\pm 1,0}, \quad  Y_{\pm 1,0}^{(j)}, \quad R_0^{(jk)}, \quad
j,k=1,\ldots,N \label{0-2}
\EEA
and has the following non-vanishing commutators \cite{hp78,negro-97,Henk97,Henkel02}
\BEA
{} \bigl[ X_n, X_{m}\bigr] &=& (n-m) X_{n+m} \;\; , \;\; \hspace{0.33truecm}
{} \bigl[ X_n, Y_m^{(j)}\bigr] \:=\: (n -m) Y_{n+m}^{(j)}
\nonumber \\
{} \bigl[ R_0^{(jk)}, Y_m^{(\ell)} \bigr] &=& \delta^{j,\ell}\,
Y_m^{(k)} - \delta^{k,\ell}\, Y_m^{(j)}
\label{gl:altcr}
\EEA
where $j,k,\ell\in\{1,\ldots,N\}$ and $n,m\in\{\pm 1,0\}$ and,
again, the $R_0^{(jk)}$ denote infinitesimal spatial rotations in
$N$ dimensions. This compact way of expressing the commutators makes
the similarities and differences with $\mathfrak{sch}(N)$ in
formulae (\ref{gl:schcr}) explicit. In particular, the subalgebra
$\mathfrak{sl}(2,\mathbb{R})=\bigl\langle X_{\pm 1,0}\bigr\rangle$
clearly appears. An explicit representation of the operators
(\ref{0-2}) is given by, summarising and generalising earlier
partial results \cite{Henkel02,Henkel05,Henkel06,Bagchi09b,mortelly-09}
\BEA X_n &=&
- t^{n+1}\partial_t - (n+1) t^n \vec{r}\cdot\vec{\nabla} -
\lambda(n+1)t^n - n(n+1) t^{n-1} \vec{\gamma}\cdot\vec{r}
\nonumber \\
Y_n^{(j)} &=& - t^{n+1} \partial_{j} - (n+1) t^n \gamma_j  \label{gl:altrep} \\
R_0^{(jk)} &=& - \bigl( r_j \partial_{k} -  r_k \partial_{j} \bigr)
- \bigl( \gamma_j \partial_{\gamma_k}-\gamma_k
\partial_{\gamma_j}\bigr); \qquad j \ne k \nonumber
\EEA
where we used the abbreviations $\partial_j = \frac{\partial}{\partial{r_j}}$
and $\partial_{\gamma_k}= \frac{\partial}{\partial{\gamma_k}}$. The
constant $\lambda$ is physically interpreted as a scaling dimension
and $\vec{\gamma}=(\gamma_1,\ldots,\gamma_N)$ is a vector of
auxiliary variables. {}From this, it can be seen that $X_{-1,0,1}$,
respectively, generate time-translations, space-time dilatations and
projective transformations, whereas $Y_n^{(j)}$ generate space
translations ($n=-1$), Galilei transformations ($n=0$) and constant
accelerations ($n=1$). We also see that although the terms
parameterised by $\gamma_j$ will create phase changes in the
transformed wave functions, they do not give rise to a central
extension, in contrast to the `mass' parameter in the Schr\"odinger
algebra, therein related to the generator $M_0$. Note the explicit
representation (\ref{gl:altrep}) of {\sc cga}$(N)$ can be extended
to a representation of an infinite-dimensional Lie algebra
\cite{hp78,Henkel02}, which might be called {\it altern-Virasoro
algebra} $\mathfrak{altv}(N) := \bigl\langle X_n, Y_{n}^{(j)},
R_0^{(jk)}\bigr\rangle$
(here  $n\in\mathbb{Z};j,k=1,\ldots N$) such that the commutation
relations (\ref{gl:altcr}) remain valid. Its central extensions were
studied in \cite{Ovsienko98}. See \cite{Duval09} for a thorough
study of the geometric interpretation of these and several other
algebras of non-relativistic space-time symmetries.

Our work is motivated by the following considerations. \\
\noindent {\bf 1.} Physicists have recently become interested in the
altern algebra in the context of non-relativistic version of the
AdS/CFT (Anti-de-Sitter/Conformal Field Theory) correspondence, see
\cite{mortelly-09,Bagchi09a} and references therein. Furthermore, it
has also been attempted to show that the equations of motions of
incompressible fluids could admit the algebra {\sc cga}$(N)$. The
current state of knowledge, discussing the various mutually
exclusive assertions, is nicely summarised  in
\cite{Horv09}. Sometimes, in order to clarify a confusing physical
situation it may be helpful to look for a precise mathematical
statement. It therefore seems appropriate to try to apply the
well-known Lie machinery \cite{ovs,ol,Fush93}
in order to construct classes of non-linear equations with  the
invariance with respect to  the altern algebra.

\noindent {\bf 2.} In $N=2$ spatial dimensions, it was recently shown in
\cite{lu-st-za-06} that the conformal Galilei algebra does admit a
so-called `exotic' central extension. This is achieved by adding to
the algebra (\ref{gl:altcr}) the following commutator \BEQ
\label{1-6} {} \bigl[ Y_n^{(1)}, Y_m^{(2)} \bigr] \:=\:
\delta_{n+m,0}\, \bigl( 3\delta_{n,0} -2 \bigr)\, \Theta, \quad
n,m\in\{\pm 1,0\}, \EEQ where the new central generator $\Theta$ is
needed for this central extension. Physicists usually call this
central extension of {\sc cga}$(2)$ the {\it exotic Galilei
conformal algebra}, and we shall denote it by {\sc ecga}. A
representation in terms of infinitesimal space-time transformations
was recently  given
in  \cite{mortelly-09}, which we repeat here
and also include the phases parameterised by the $\gamma_j$:
\BEA
{} X_n &=& -t^{n+1}\partial_t - (n+1) t^n \vec{r}\cdot\vec{\nabla} -
\lambda(n+1) t^n  - (n+1)n t^{n-1} \vec{\gamma}\cdot\vec{r} - (n+1)n
\vec{h}\cdot\vec{r}
\nonumber \\
Y_n^{(j)} &=& - t^{n+1}\partial_{j} - (n+1) t^n \gamma_j - (n+1) t^n
h_j - (n+1) n (r_2-r_1) \theta \label{gl:exaltrep} \\
R_0^{(12)} &=& - \bigl( r_1 \partial_{2} -  r_2 \partial_{1} \bigr)
- \bigl( \gamma_1 \partial_{\gamma_2}  -
\gamma_2\partial_{\gamma_1}\bigr) - \frac{1}{2\theta} \vec{h}\cdot
\vec{h} \nonumber
\EEA
where $n\in\{\pm 1,0\}$ and $j,k\in\{1,2\}.$
Because of Schur's lemma, the central generator $\Theta$ can be
replaced by its eigenvalue $\theta\ne 0$. The components of the
vector-operator $\vec{h}=(h_1,h_2)$ are connected by   the
commutator $[h_1, h_2] = \Theta$. We are interested in finding
systems of non-linear PDEs which are invariant under the exotic
conformal Galilei algebra  {\sc ecga} and its subalgebras.

The paper is organised as follows. In section~2, we use the known
techniques for finding Lie symmetries \cite{ovs, ol, Fush93} and
conditional symmetries \cite{Fush93} to construct non-linear partial
differential equations (PDEs) with the  {\sc cga}$(N)$-symmetry,
although explicit results will only be written down for $N=2$ and
shown how they extend to the case $N>2$. Since the notion of
conditional symmetry requires the introduction of an auxiliary
condition, we also look for {\em pairs} of non-linear equations
which are invariant under the Lie algebra {\sc cga}$(2)$. In
section~3, we extend these results to the case of {\sc ecga}. Large
classes of systems of non-linear PDEs, which are invariant under
{\sc ecga} and some of its subalgebras, are found. In section~4, we
illustrate the results obtained  through examples of
correctly-specified systems, which can be reduced to the
hydrodynamic equations of motion of a two-dimensional incompressible
fluid subject to certain forces. Our conclusions are given in
section~5.

\section{Conformal Galilei-invariance and nonlinear PDEs }

In this section, we use the standard representation
(\ref{gl:altrep}) for the space-time transformations in the
conformal Galilei algebra algebra {\sc cga}$(N)$, but normalise it
such that $\lambda=0$ and $\gamma_j=0$. For the sake of simplicity,
and since we plan to consider the exotic central extension later, we
restrict our attention to $N=2$ from now on. Thus, we consider here
the following representation of the algebra {\sc cga}$(2)$
\BEQ \label{2-3}
X_n = - t^{n+1}\partial_t - (n+1) t^n \bigl( r_1\partial_1 + r_2\partial_2\bigr), \;\;
Y_n^{(1)} = - t^{n+1}\partial_{1}, \;\;
Y_n^{(2)} = - t^{n+1} \partial_{2}, \;\;
R_0^{(12)} = - \bigl( r_1\partial_{2} -  r_2 \partial_{1} \bigr),
\EEQ
where  $n=\pm 1,0$. Our aim is  to describe all second-order
PDEs, which are invariant under the altern algebra  in the
(1+2)-dimensional space of independent variables $t, r_1, r_2$. The
most general form of such a PDE reads
\be\label{2-4}
H\left(t, r_1, r_2,u,\underset{1}{u},\underset{11}{u}\right) =0
\ee
Here and afterwards, we use the notations
$\underset{1}{u}=(u_t,u_1,u_2)=(\partial u/\partial t,
\partial u/\partial r_1, \partial u/\partial r_2)$,
$\underset{11}{u}=(u_{tt},u_{t1},\ldots,u_{22})$,
$\Delta=\partial^2_1 + \partial^2_2$ is the spatial Laplacian and
$H$ is an arbitrary smooth function.

First of all, we consider the subalgebra with basic operators
$\bigl\langle X_{-1}, Y_{-1,0}^{(j)}, R^{(12)}\bigr\rangle$, which
is well-known Galilei algebra with zero mass. Hereafter, we use the
notation  $\mathfrak{gal}^{(0)}(N)$ with $N=2$ for this algebra for
which the notation $AG_0(1.N)$ is also common, see \cite{Cher01}.
Our starting point is the following well-established result
\cite[Theorem 7]{Cher01}: {\it an arbitrary second-order PDE
(\ref{2-4}) is invariant under the massless Galilei algebra
$\mathfrak{gal}^{(0)}(2)$ if and only if it can be written in the
form} \be\label{2-5} H_{\mathfrak{gal}} \left(u, \, u_a u_a,\,
\Delta u, u_au_bu_{ab}, u_{11}u_{22}-u_{12}^2, W^{I}, W^{II} \right)
=0, \ee {\it where  the summation convention over repeated indices
$a=1,2$ and $b=1,2$ is implied and  $H_{\mathfrak{gal}}$  is an
arbitrary smooth function and} \be\label{2-6} W^I := \det\left[
\begin{array}{ccc} u_t & u_1 & u_2 \\ u_{t1} & u_{11} & u_{12} \\ u_{t2} & u_{12} &  u_{22}
\end{array} \right]
\;\; , \;\;
W^{II} := \det\left[
\begin{array}{ccc} u_{tt} & u_{t1} & u_{t2} \\ u_{t1} & u_{11} & u_{12} \\ u_{t2} & u_{12} & u_{22}
\end{array} \right].
\ee

In other words, the seven arguments of the function
$H_{\mathfrak{gal}}$ form a full set of absolute differential
invariants (up to the second order) of the Galilei algebra
$\mathfrak{gal}^{(0)}(2)$. Recall that the invariants
$u_{11}u_{22}-u_{12}^2$ and  $W^{II}$ are nothing else but the
right-hand-side (RHS) of the well-known Monge-Amp\`ere equation in
two- and three-dimensional space, respectively \cite{Pogo75}, while
the invariant determinant $W^{I}$ was derived for the first time in
\cite{Fush85}.

Now consider the generators $Y_1^{(1)}$ and $Y_1^{(2)}$  of constant
accelerations, which lead to the finite transformations
\be\label{2-8}
t' =t \;\; , \;\; r_a' = r_a + v_a t^{2} \;\; , \;\; u' = u,
\ee
where $v_1$ and $v_2$ are arbitrary parameters. It can
be checked by direct computation that a PDE belonging to  the  class
(\ref{2-5}) is invariant under the transformations (\ref{2-8}) only
if the function $H_{\mathfrak{gal}}$ does not depend on $W^{II}$.
Next, consider the scale-transformations generated by the operator
$X_0=-t\partial_t - r_j\partial_j$ one obtains in a similar way a
further restriction on the function $H_{\mathfrak{gal}}$. Therefore,
we arrive at our first result,
where we consider the subalgebra of {\sc cga}$(2)$ which is obtained when leaving out
the accelerations $Y_1^{(j)}, \ j=1,2$ and projective transformations $X_1$.

\bt A PDE of the form (\ref{2-5}) is invariant under the following
subalgebra of the conformal Galilei algebra
\BD
\bigl\langle
X_{-1,0},Y_{-1,0}^{(j)},R_0^{(12)}\bigr\rangle_{j=1,2}\subset\mbox{\sc cga}(2)
\ED
with the realisation (\ref{2-3}),
if and only if  this equation possesses the form
\be\label{2-9}
H_{0}\left(u, \, Z_1,\, Z_2, \,Z_3,\, Z_4 \right) =0,
\ee
where $H_{0}$  is an arbitrary smooth function and
\BEA
Z_1 := {\Delta u}\cdot\bigl({u_au_a}\bigr)^{-1} &,&
Z_2 := {\bigl( u_au_bu_{ab}\bigr)}\cdot{\bigl(u_au_a\bigr)^{-2}} \;\; ,
\nonumber \\
Z_3 := {\bigl( u_{11}u_{22}-u_{12}^2\bigr)}\cdot{\bigl(u_au_a\bigr)^{-2}} &,&
Z_4 := {W^{I}}\cdot{\bigl(u_au_a\bigr)^{-5/2}}. \label{2-10} \
\EEA
\et

\noindent \textbf{Remark 1.} Theorem 1 can be easily generalised to
the case $N=3$ using paper \cite{Cher01}. Generalisation to higher
dimensionality is also possible, but will lead to very lengthy
expressions.

One observes that the invariant $Z_4$ is the only one in equation
(\ref{2-9}) which contains time derivatives. It follows, and can
also be checked directly, that $Z_4$ cannot be modified by the
invariants $u, \, Z_1,\, Z_2$ and $ Z_3$ to a form that is invariant
under projective transformations
\be\label{2-11}
t' = \frac{t}{(1-pt)} \;\; , \;\; r_a' = \frac{r_a}{(1-pt)^2}  \;\; , \;\;  u' = u,
\ee
generated by the operator $X_1$ from (\ref{2-3}) with
$N=2$ (hereafter $p$ is the group parameter). In other words, a
given PDE belonging to the class (\ref{2-9}) can be invariant under
the operator $X_1$ only in the case when it does not contain time
derivatives, i.e., it contains the variable $t$ as a parameter. We
are not interested in this case and hereafter consider only PDEs
containing time derivatives.

Thus we can conclude: {\it there are no non-trivial PDEs
belonging to the class (\ref{2-4}) that are invariant under the representation (\ref{2-3})
of the conformal Galilei algebra {\sc cga}$(2)$.}

Indeed, we found earlier an analogous result for the
infinite-dimensional extension of the representation (\ref{2-3})
where $n\in\mathbb{Z}$ \cite{ch-he-2004}.  Following the same lines
as in \cite{ch-he-2004}, we use now the concept of conditional
invariance to construct PDEs that are {\it conditionally invariant}
under {\sc cga}$(2)$. The notion of conditional invariance was
introduced in \cite{Fush88} as a generalisation of the well-known
non-classical symmetry \cite{Blum69}.

\noindent {\bf Definition.} \cite [Section 5.7]{Fush93} {\it A PDE
of the form} \BEQ \label{2:gl:c1}
S\left(t,\vec{r},u,\underset{1}{u},\underset{11}{u}\right) =0 \EEQ
{\it is said to be {\em conditionally invariant} under the operator}
\BEQ \label{2:gl:Q} Q = \xi^0(t,\vec{r},u)\partial_t + \xi^a
(t,\vec{r}, u)\partial_a + \eta(t,\vec{r},u)\partial_u \EEQ {\it
where $\vec{r}=(r_1,\ldots,r_N)$ and  $\xi^a$ with $a=0,1,\ldots,N$
and $\eta$  are smooth functions, if it is invariant (in Lie's
sense) under this operator only together with an additional
condition of the form} \BEQ \label{2:gl:c2}
S_Q\left(t,\vec{r},u,\underset{1}{u},\underset{11}{u}\right) =0 \EEQ
{\it that is, the over-determined system of equations
(\ref{2:gl:c1},\ref{2:gl:c2}) is invariant under a Lie group
generated by the operator $Q$.}

With this concept at hand, we can now prove the following.

\bt A PDE of the form (\ref{2-9}) is conditionally invariant under
the algebra {\sc cga}$(2)$ with the realisation
(\ref{2-3}) if and only if the
condition holds:
\be\label{2-12}
W^{III} := \det\left[
\begin{array}{ccc} 0   & u_{1} &u_{2} \\ u_{1} & u_{11} & u_{12} \\ u_{2} & u_{12} & u_{22}
\end{array} \right]=0.
\ee
\et
\noindent \textbf{Proof.}
One needs to establish a necessary and
sufficient condition when a system, consisting of an arbitrary  PDE of
the form (\ref{2-9}) and the condition in question, is invariant
under the projective transformations (\ref{2-11}). It is a
straightforward calculation that the transformations (\ref{2-11})
transform the derivatives as follows:
\be\label{2-12a}
\begin{array}{cc}
u'_{a'} = u_{a}(1-pt)^2, \ a=1,2 ,  & u'_{t'} = u_{t}(1-pt)^2-2p r_a u_a(1-pt),\\
u'_{a'b'} = u_{ab}(1-pt)^4, \ a,b=1,2 , &  \qquad
u'_{t'a'} = (1-pt)^3\Bigl((1-pt)u_{ta}-2p r_b u_{ab} - 2p u_{a}\Bigr).
\end{array}
\ee
So, using formulas (\ref{2-10}) and (\ref{2-12a}) one easily
shows that
\be\label{2-12b}
Z^{'}_1=Z_1, \quad Z^{'}_2=Z_2, \quad Z^{'}_3=Z_3,
\ee
i.e., all arguments of the function $H_0$ from eq.(\ref{2-9}), excepting $Z_4$, are invariant under transformations
(\ref{2-11}). It turns out,  $Z_4$ is not invariant with respect to
the projective transformations. In fact, substituting  formulas
(\ref{2-12a}) into the expression
\be\label{2-12c}
Z^{'}_4 :={(W^{I})'}\cdot{\bigl(u'_{a'}u'_{a'}\bigr)^{-5/2}},
\ee
we obtain
\be\label{2-12d}
Z^{'}_4= Z_4-2p(1-pt)^{-1}W^{III}.
\ee
Thus, the projective transformations (\ref{2-11}) bring the PDE (\ref{2-9}) to the
form
\be\label{2-12e}
H_{0}\left(u', \,Z^{'}_1 ,\, Z^{'}_2, \,Z^{'}_3,\, Z^{'}_4+2p(1-pt)^{-1}W^{III} \right) =0.
\ee
Since  $p$ is an arbitrary parameter, we conclude that expression (\ref{2-12e})
must take the form
\be\label{2-12f}
H_{0}\left(u', \,Z^{'}_1 ,\, Z^{'}_2, \,Z^{'}_3,\, Z^{'}_4 \right) =0
\ee
if and only if the condition (\ref{2-12}) holds. Hence, the PDE (\ref{2-9}) is conditionally
invariant under the algebra {\sc cga}$(2)$ and the
corresponding condition must be equation (\ref{2-12}). This
completes the proof. \hfill $\blacksquare$

\noindent \textbf{Remark 2.} Condition (\ref{2-12}) contains the
time variable as a parameter and can be rewritten in  the equivalent
form $ u_au_a\Delta u = u_au_bu_{ab}$, i.e., $Z_1=Z_2$.

\noindent \textbf{Remark 3.} The simplest  PDE that is conditionally
invariant under {\sc cga}$(2)$ is $W^{I}=0$. Lie symmetries and
a wide range of exact solutions for this equation have been
constructed in \cite{Fush85,ch-87}.


\noindent
\textbf{Remark 4.} Theorem 2 can be  generalised to the case $N=3$.
The condition needed has the form
\[ W^{III} := \det\left[
\begin{array}{cccc}
 0     & u_{1}  & u_{2}  & u_{3}  \\
 u_{1} & u_{11} & u_{12} & u_{13} \\
 u_{2} & u_{12} & u_{22} & u_{23} \\
 u_{3} & u_{31} & u_{32} & u_{33}
\end{array} \right]=0.
\]

Theorems 1 and 2 state that although there are no second-order PDEs
admitting   {\sc cga}$(2)$-symmetry in the Lie sense, there is
a wide class of such PDEs possessing this symmetry algebra when the
concept of conditional invariance is used. In order to understand
where this result comes from, we consider now the {\it system} of
PDEs
\be\label{2-14}
\begin{array}{ccc} H_1\left(t,
r_1, r_2,u,v,\underset{1}{u},\underset{11}{u},
\underset{1}{v},\underset{11}{v}\right) &=&0 \\
H_2\left(t, r_1, r_2,u,v,\underset{1}{u},\underset{11}{u},
\underset{1}{v},\underset{11}{v}\right) &=&0,
\end{array}
\ee
where $H_1$  and $H_2$ are arbitrary smooth functions while $u(t,r_1, r_2)$ and
$v(t, r_1, r_2)$ are unknown functions.

\bt A system of PDEs of the form (\ref{2-14}) is invariant under the
algebra {\sc cga}$(2)$ with the realization (\ref{2-3})
if it takes the form
\be\label{2-15}
\begin{array}{ccc} H_{CGA}\left(u, \, v, \, Z^{uv},
\, Z^u_1,\ldots,Z^u_4, \,Z^v_1,\ldots,Z^v_4, \right) &=&0 \\
\lambda_1W^{III}(u,u,u)+\lambda_2W^{III}(v,u,u)+\lambda_3W^{III}(u,v,v)+\lambda_4W^{III}(v,v,v)
&=&0  \end{array} \ee where $H_{CGA}$ is an arbitrary smooth
function, $\lambda_1,\ldots, \lambda_4$ are arbitrary constants (at
least one of them must be non-zero) and the notations are used:
\be\label{2-16}
\begin{array}{cccc} Z^u_1=& {u_au_a}/{u_av_a}, \, &
Z^v_1=& {v_av_a}/{u_av_a}\\
Z^u_2= &{\Delta u}/{u_av_a}, \, & Z^v_2= & {\Delta v}/{u_av_a}\\
Z^u_3=& {\bigl(u_au_bu_{ab}\bigr)}/{\bigl(u_av_a\bigr)^2}, \,& Z^v_3=& {\bigl(v_av_bv_{ab}\bigr)}/{\bigl(u_av_a\bigr)^2}\\
Z^u_4=& {\bigl(u_{11}u_{22}-u_{12}^2\bigr)}/{\bigl(u_av_a\bigr)^2}, \,& Z^v_4=&
{\bigl(v_{11}v_{22}-v_{12}^2\bigr)}/{\bigl(u_av_a\bigr)^2},
\end{array}
\ee
\be\label{2-17}
Z^{uv}= (u_au_a)^{-5/2}\Bigl(\lambda_1W^{I}(u,u,u)+
\lambda_2W^{I}(v,u,u)+\lambda_3W^{I}(u,v,v)+\lambda_4W^{I}(v,v,v)\Bigr);
\ee
\be\label{2-20}
W^I(u,v,w) := \det\left[
\begin{array}{ccc} u_t & u_1 & u_2 \\ v_{t1} & v_{11}  & v_{12} \\ w_{t2} & w_{12} &  w_{22}
\end{array} \right], \quad
W^{III}(u,v,w) := \det\left[ \begin{array}{ccc} 0   &
u_{1} &u_{2} \\ v_{1} & v_{11}  & v_{12} \\
 w_{2} & w_{12} &  w_{22}
\end{array} \right].
\ee
\et

\medskip

\noindent \textbf{Proof:} This result is readily derived by direct
application of the continuous transformations generated by the basic
operators (\ref{2-3}) to the system (\ref{2-15}). \hfill
$\blacksquare$

The system (\ref{2-15}) contains the time derivatives only in the
case if the function  $H_{CGA}$ explicitly depends on $Z^{uv}$. The
second equation of the system contains the variable $t$ as a
parameter and reproduces the condition (\ref{2-12}) if one sets
$\lambda_1=1, \lambda_2=\lambda_3=\lambda_4=0$ (simultaneously
$Z^{uv}$ is reduced to $Z_4$).

\section{Systems of non-linear PDEs  invariant under the exotic conformal Galilei algebra}

In this section we study possibilities to construct systems of PDEs
that are invariant under {\sc ecga}. This can only be done in $N=2$
space dimensions, since only then the exotic central extension
exists \cite{lu-st-za-06}. The application of the Lie scheme
requires an explicit realization in terms of linear first-order
differential operators, which have  been constructed very recently
\cite{mortelly-09} and  listed in (\ref{gl:exaltrep}). It means that
the operators $h_1, h_2$ and  $\Theta$ must be correctly specified.
According to the paper \cite{mortelly-09},  the possible explicit
realization of the algebra {\sc ecga} reads as
\BEA \label{3-1}
X_{-1}&:=&- \partial_t \;\; , \;\; Y_{-1}^{(1)}:= -\partial_1 \;\; ,
\;\; Y_{-1}^{(2)}:=-\partial_2 \;\; , \;\;
\Theta: =\partial_w \\
\label{3-2}  Y_0^{(1)}&:=& -t\partial_1+\partial_{v^1}
-\frac{v^2}{2}\partial_w,
\hspace{2.3truecm},
 \;\;
Y_0^{(2)} :=-t\partial_2+ \partial_{v^2} +\frac{v^1}{2}\partial_w \\
\label{3-3}  Y_1^{(1)} &:=&-t^2\partial_1 + 2t\partial_{v^1} -
(2r_2+tv^2)\partial_w \hspace{0.5truecm}, \;\;
 Y_1^{(2)}:= -t^2\partial_2 +2t\partial_{v^2} + (2r_1+tv^1)\partial_w \\
\label{3-4}
 X_0&:=&- t\partial_t - r_1\partial_1 - r_2\partial_2 \\
\label{3-5} X_1&:=& -t\bigl(t\partial_t + 2r_1\partial_1+
2r_2\partial_2\bigr)
+r_1\partial_{v^1}+r_2\partial_{v^2}-(r_1v^2-r_2v^1)\partial_w \\
\label{3-6}
 R_0^{(12)}&:=&- r_1\partial_2 + r_2\partial_1 - v^1\partial_{v^2} + v^2\partial_{v^1},
\EEA where $\partial_w=\partial/\partial {w}$,
$\partial_{v^1}=\partial/\partial {v^1}$,
$\partial_{v^2}=\partial/\partial {v^2}$.

Clearly, one has the massless Galilei subalgebra with the basic
operators $\mathfrak{gal}^{(0)}(2)=\bigl\langle
 X_{-1}, Y_{-1,0}^{(j)}, R^{(12)}\bigr\rangle$.
Because of the additional variables $v^1, v^2$ and $w$, however,
this realisation cannot be reduced to the one studied in sections~1
and~2. The case when the generators $Y_{0}^{(j)}$ of Galilei
transformations depend only on the variables $v^1$ and  $v^2$ (set
formally $\partial_w=0$ in (\ref{3-2})) is rather typical
\cite{Cher01}.

To prepare for the construction of  PDEs that admit {\sc ecga} with
the realisation (\ref{3-1}--\ref{3-6}) one should choose dependent
and independent variables. The example given in \cite{mortelly-09}
suggests that all the variables $t, r_1, r_2, v^1, v^2$ and $w$ are
independent. Furthermore, a linear second-order PDE on the function
$\Psi(t, r_1, r_2, v^1, v^2,w)$, which is invariant under {\sc
ecga}, is constructed. That example is very simple but rather
artificial. In fact,  the equation (3.11) \cite{mortelly-09} reads
as \be\label{3-*} \frac{\partial^2 \Psi}{\partial t\partial w} -
\frac{\partial^2 \Psi}{\partial r_1 \partial v^2}+\frac{\partial^2
\Psi}{\partial r_2 \partial v^1} - \frac{v^1}{2}\frac{\partial^2
\Psi}{\partial r_1 \partial w} - \frac{v^2}{2}\frac{\partial^2
\Psi}{\partial r_2 \partial w}=0. \ee Using Maple,  we have
established that this equation is invariant under an 28-dimensional
Lie algebra containing the 11-dimensional algebra {\sc ecga} with
realisation (\ref{3-1}--\ref{3-6}) as a subalgebra. Since the
standard 6-dimensional wave equation is also invariant under this
28-dimensional Lie algebra, which is the conformal algebra with the
standard realization, equation (\ref{3-*}) can be nothing else but
the wave equation in an unusual form.

Our aim is to construct PDEs in the (1+2)-dimensional space of the
variables $t, r_1, r_2$. This immediately leads to the need to
consider {\it systems of PDEs}  with the unknown functions $v^1,
v^2$ and $w$. Hereafter the notations \be\label{3-7}
 u^1=v^1, \quad u^2=v^2, \quad u^3=w
\ee are used to simplify the corresponding formul{\ae}.

We start from the most general system of first-order  PDEs
\be\label{3-8} B^k=0 \;\; , \;\; k=1,2,3 \ee
 where $B^1, B^2$ and
$B^3$ are arbitrary sufficiently smooth functions  independent
variables $t, r_1, r_2$, dependent variables (\ref{3-7}) and their
first-order derivatives.
Obviously, to construct systems with the {\sc ecga}-invariance, one
straightforwardly obtains that all the functions in (\ref{3-8})
cannot depend on the variables $t, r_1, r_2$ and $u^3$ (see the
generators (\ref{3-1}) of the exotic conformal Galilei algebra).
Hence $B^1, B^2$ and $B^3$ are assumed to be functions on
\be\label{3-11} u^1, \, u^2, \, \underset{1}{u^1},\,
\underset{1}{u^2}, \, \underset{1}{u^3}, \ee where
$\underset{1}{u^k}= (u^k_0,u^k_1,u^k_2), \, k=1,2,3$. To avoid
possible misunderstanding, we also assume that
 system (\ref{3-8}) doesn't contain algebraic
equation(s) and cannot be reduced to one containing algebraic
equation(s).

It will be helpful to proceed step by step and to consider several
subalgebras of the algebra {\sc ecga}. Specifically, we shall
consider the following cases:
\begin{enumerate}
\item the 8-dimensional subalgebra $\mathfrak{ea}_1 :=
\bigl\langle X_{-1}, Y_{-1,0,1}^{(j)},
\Theta \bigr\rangle \, j=1,2$, which contains the operators of time-
and space-translations, the Galilei transformations, accelerations
and the generator $\Theta$ for the central extension. The Lie
algebra $\mathfrak{ea}_1$
is semidirect sum of the Weil algebra and the Abelian algebra
generated by  the    two 'exotic' operators (\ref{3-3}) and the
operator $\Theta$.
\item the 10-dimensional subalgebra $\mathfrak{ea}_2 :=
\bigl\langle
X_{-1,0,1}, Y_{-1,0,1}^{(j)},\Theta \bigr\rangle$, where with
respect to $\mathfrak{ea}_1$ the dilatation operator $X_0$ and the
projective operator  $X_1$ are added.
\item the 10-dimensional subalgebra $\mathfrak{ea}_3 := \bigl\langle
X_{-1,0}, Y_{-1,0,1}^{(j)},\Theta, R_0^{(12)} \bigr\rangle$, where
with respect to $\mathfrak{ea}_1$ the dilatation operator $X_0$ and
the rotation operator $R_0^{(12)}$  are added.
\item the 11-dimensional  {\sc ecga} with the basic operators
(\ref{3-1}--\ref{3-6}).
\end{enumerate}

\bt A system of PDEs system  of the form (\ref{3-8}) is invariant
under the Lie algebra $\mathfrak{ea}_1$ if and only if it is of the
form \be\label{3-12} b^k\Big(u^1_1, u^1_2, u^2_1, u^2_2, W_1,
W_2,W_3\Big)=0,\quad k=1,2,3 \ee
 where $b^1$, $b^2$
and $b^3$ are arbitrary smooth functions of the variables $u^1_1,
u^1_2, u^2_1, u^2_2, W_1, W_2$ and $W_3$. Here, the notations
\be\label{3-15} W_1:= 2u^1_0 + 2u^3_2 - 2u^1u^1_1 + u^1u^2_2 -
3u^2u^1_2, \ee \be\label{3-16} W_2:= 2u^2_0 - 2u^3_1 - 2u^2u^2_2 +
u^2u^1_1 - 3u^1u^2_1, \ee \be\label{3-17} W_3:= 2u^3_0 - u^2u^1_0 +
u^1u^2_0 -2u^1u^3_1 - 2u^2u^3_2
 + u^1u^2u^1_1 - u^1u^2u^2_2 - u^1u^1u^2_1 + u^2u^2u^1_2
\ee are used. \et

\noindent \textbf{Proof.} Consider the algebra $\mathfrak{ea}_1$. A
straightforward analysis
 shows that  system
(\ref{3-8}) will automatically be invariant under the Galilei
operators (\ref{3-2}) if it is invariant with respect to the
acceleration operators (\ref{3-3}). Thus, to construct systems with
$\mathfrak{ea}_1$-symmetry
we need to find among systems of the form (\ref{3-8}) those, which
are invariant under the operators $Y_{1}^{(j)}, \, j=1,2$, i.e.
under the operator \be\label{3-40} X \equiv -\alpha_1Y_{1}^{(1)} -
\alpha_1Y_{1}^{(1)}= \alpha_1t^2\partial_1 + \alpha_2t^2\partial_2
-2\alpha_1t\partial_{u^1} - 2\alpha_2t\partial_{u^2}
+\Big(\alpha_1(2r_2+tu^2)-\alpha_2(2r_1+tu^1)\Big)\partial_{u^3},
\ee where $\alpha_1$ and $\alpha_2$ are arbitrary parameters.

 The construction is based on  a direct
application of the classical Lie algorithm \cite{ovs,ol, Fush93}.
According to this algorithm, we consider system (\ref{3-8}) as the
manifold $(S_1, S_2, S_3)$ determined by the restrictions:
\be\label{3-41} S_k\equiv B^k\Big(u^1, \, u^2, \,
\underset{1}{u^1},\, \underset{1}{u^2}, \, \underset{1}{u^3}\Big) =
0 \;\; , \;\; k=1,2,3 \ee
 in the space of the variables
 \[t,\, r_1,\, r_2,\, u^1, \, u^2, \,u^3,  \, \underset{1}{u^1},\,
\underset{1}{u^2}, \, \underset{1}{u^3}. \] In order to determine
these  unknown functions $B^k$ one needs to use the invariance
conditions
\be\label{3-42}  
\bar XB^k \vert_{S_j=0, \, j=1,2,3}=0, \;\; k=1,2,3, \ee where $\bar
X$ is the first prolongation of the operator $X$ (\ref{3-40}). The
explicit form of $\bar X$ is  calculated by the well-known
prolongation formulae (see, e.g. \cite{ovs,ol, Fush93}). In the case
of of the operator $X$ (\ref{3-40}), these formulae lead to the
operator
 \be\label{3-43}\bar X = X - 2(\alpha_1+\alpha_1 tu_1^1 +\alpha_2
tu_2^1)\partial_{u_0^1} - 2(\alpha_2+\alpha_2 tu_2^2 +\alpha_1
tu_1^2)\partial_{u_0^2}+ \ee \[+\Big(\alpha_1u^2-\alpha_2u^1
+\alpha_1 t(u_0^2-2u_1^3) -\alpha_2
t(u_0^1+2u_2^3)\Big)\partial_{u_0^3}+ (-2\alpha_2-\alpha_2 tu_1^1
+\alpha_1 tu_1^2)\partial_{u_1^3}+ (2\alpha_1-\alpha_2 tu_2^1
+\alpha_1 tu_2^2)\partial_{u_2^3}.\]

Substituting  operator (\ref{3-43}) into system (\ref{3-42}) and
carrying out  the relevant
 calculations, we obtain
 a system of linear first-order PDEs  for finding  the functions
 $B^k$. Firstly, we aim to find the set
 of {\it absolute zero- and
first-order differential invariants} of the algebra
$\mathfrak{ea}_1$ and then to show that the functions $B^k$ can
depend only on those invariants. To construct the set mentioned
above one needs to
 substitute operator (\ref{3-43})
into system (\ref{3-42}) without the conditions $S_j=0, \, j=1,2,3,$
and to solve  three linear first-order PDEs obtained (they are
omitted here). Since those PDEs contain two arbitrary parameters and
the unknown functions don't depend on the time-variable, each PDE
can be splitted with respect to the $\alpha_1, \alpha_2, \alpha_1 t$
and $\alpha_2 t$. Thus, we arrive at the system of 12 PDEs:
\be\label{3-44} -2\partial_{u_0^1}B^k +u^2\partial_{u_0^3}B^k
+2\partial_{u_2^3}B^k = 0 \ee \be\label{3-45} 2\partial_{u_0^2}B^k
+u^1\partial_{u_0^3}B^k +2\partial_{u_1^3}B^k = 0 \ee
\be\label{3-46} 2\partial_{u^1}B^k
+2u_1^1\partial_{u_0^1}B^k+2u_1^2\partial_{u_0^2}B^k+
(2u_1^3-u_0^2)\partial_{u_0^3}B^k -u_1^2\partial_{u_1^3}B^k -
u_2^2\partial_{u_2^3}B^k= 0 \ee \be\label{3-47} 2\partial_{u^2}B^k
+2u_2^1\partial_{u_0^1}B^k+2u_2^2\partial_{u_0^2}B^k+
(2u_2^3+u_0^1)\partial_{u_0^3}B^k +u_1^1\partial_{u_1^3}B^k
+u_2^1\partial_{u_2^3}B^k= 0, \ee where $k=1,2,3$. Now one notes
that this system  consists of three separate subsystems for $k=1,
k=2$ and $k=3$ so that its  general solution can be found by solving
these subsystems. A straightforward application of the standard
technics leads to the following general solution:
 \be\label{3-48} B^k= b^k\Big(u^1_1, u^1_2, u^2_1, u^2_2, W_1,
W_2,W_3\Big), \ee where $W_j, j=1,2,3$ are defined in
(\ref{3-15}--\ref{3-17}). Thus, there are  exactly seven absolute
first-order  differential invariants (no zero-order invariants!) of
the algebra $\mathfrak{ea}_1$ so that an arbitrary system of the
form (\ref{3-12}) admits this algebra.

To prove that system (\ref{3-12}) is the most general system with
$\mathfrak{ea}_1$-symmetry, one needs to solve
(\ref{3-42}--\ref{3-43}), i.e. to take into account the conditions
$S_j=0, \, j=1,2,3$. We assumed from the very beginning  that system
(\ref{3-8}) doesn't contain algebraic equation(s) and $B^k$ are
arbitrary sufficiently smooth functions, hence system (\ref{3-8})
can be solved with respect to three different first-order
derivatives. Assuming that these derivatives are $u_0^1, u_0^2$ and
$u_0^3$, equations (\ref{3-41}) take the form \be\label{3-49}
S_k\equiv D^k\Big(u^1, \, u^2, \, u_1^1,\, u_1^2, \, u_1^3,\,
u_2^1,\, u_2^2, \, u_2^3 \Big)- u_0^k= 0 \;\; , \;\; k=1,2,3 \ee
where $D^k$  are   smooth functions, which can be assumed arbitrary.
Substituting operator (\ref{3-43}) into system (\ref{3-42}),
applying (\ref{3-49}) to eliminate the variables $u_0^1, u_0^2$ and
$u_0^3$ and carrying out the relevant
 calculations, we obtain
 a system of linear first-order PDEs  for finding  the functions
 $D^k$. Each PDE
can be again  splitted with respect to the $\alpha_1, \alpha_2,
\alpha_1 t$ and $\alpha_2 t$. Finally, we arrive at the system of 12
PDEs, which is nothing else but system (\ref{3-44})-(\ref{3-47})
with \be\label{3-50} B^k = D^k\Big(u^1, \, u^2, \, u_1^1,\, u_1^2,
\, u_1^3,\, u_2^1,\, u_2^2, \, u_2^3 \Big)- u_0^k, \, k=1,2,3.\ee
Setting $k=1$ one obtains the subsystem \be\label{3-51}
\partial_{u_2^3}D^1=-1, \quad \partial_{u_1^3}D^1= 0 \ee
\be\label{3-52} 2\partial_{u^1}D^1=2u_1^1-u_2^2, \quad
2\partial_{u^2}D^1=3u_2^1 \ee to find the function $D^1$. Solution
of this system is rather trivial and gives \be\label{3-53}
D^1=\frac{1}{2}d^1(u_1^1,\, u_1^2, \, u_2^1,\,
u_2^2)-\frac{1}{2}(2u^3_2 - 2u^1u^1_1 + u^1u^2_2 - 3u^2u^1_2),\ee
where $d^1$ is an arbitrary   function. Substituting $D^1$ into
(\ref{3-49}) for $k=1$, we immediately obtain that the first
equation of system (\ref{3-8}) must possess the form \be\label{3-54}
W_1 = d^1(u_1^1,\, u_1^2, \, u_2^1,\, u_2^2). \ee In a quite similar
way, the second and third equation were found: \be\label{3-55} W_2 =
d^2(u_1^1,\, u_1^2, \, u_2^1,\, u_2^2), \quad  W_3 = d^3(u_1^1,\,
u_1^2, \, u_2^1,\, u_2^2).\ee However, it is easily seen that system
(\ref{3-54}-\ref{3-55}) is a particular case of the
$\mathfrak{ea}_1$-invariant system (\ref{3-12}).

If system (\ref{3-8}) cannot  be solved with respect to the
variables $u_0^1, u_0^2$ and $u_0^3$ then one must be solved with
respect to   another triplet of derivatives. Nevertheless, making
the relevant calculations,  a system of 12 PDEs  is obtained, which
will be again a particular case of  system
(\ref{3-44})-(\ref{3-47}). Thus, its solution will lead to another
particular case of system (\ref{3-12}).This completes the proof.

\hfill $\blacksquare$

\noindent \textbf{Remark 5.}  Formul{\ae} (\ref{3-15}--\ref{3-17})
and $u^1_1, u^1_2, u^2_1, u^2_2$ present the full  set of absolute
first-order differential invariants of the algebra
$\mathfrak{ea}_1$.

\bt A system of PDEs  of the form (\ref{3-12}) is invariant under
the  Lie algebra $\mathfrak{ea}_2$ if and only if it can be written
in the form \be\label{3-25} h^k\Big(U^1, U^2, W^*_1,
W^*_2,W^*_3\Big)=0,\quad k=1,2,3, \ee where $h^1$, $h^2$ and $h^3$
are arbitrary smooth functions of five variables \be\label{3-28}
W^*_k:=\frac{W_k}{u^1_2-u^2_1}, \, k=1,2,3, \quad
\frac{u^1_1-u^2_2}{u^1_2-u^2_1}, \quad
\frac{u^1_2}{u^1_2-u^2_1}. \ee \et

\noindent \textbf{Proof:} To construct all PDEs' systems with
$\mathfrak{ea}_2$-invariance, we need to find among systems of the
form (\ref{3-12}) those,  which are invariant under the operator
(\ref{3-5}). Note that scale-invariance (see operator (\ref{3-4}))
will automatically be obtained if the invariance with respect to the
projective  operator (\ref{3-5}) holds true.

The operator (\ref{3-5}) produces the projective  transformations
\be\label{3-20}
t \mapsto t' = \frac{t}{(1-p t)} \;\; , \;\;
r_a \mapsto r_a' = \frac{r_a}{(1-p t)^2}
\ee
for the independent variables and
\be\label{3-21}
u^a \mapsto (u^a)' = u^a - \frac{2p r_a}{1-p t}, \;\; , a=1,2, \quad
u^3 \mapsto (u^3)' = u^3 -
\frac{p(r_1 u^2-r_2 u^1)}{1-p t}
\ee
for the dependent variables.

One may directly check that  the transformations
(\ref{3-20}--\ref{3-21}) transform the set of absolute first-order
differential invariants of the algebra $\mathfrak{ea}_1$ as follows:
\be\label{3-29} u^a_a \mapsto (u^a)'_{a'} = (1-p t)^2u_a^a -
2p(1-pt), \quad u^a_b \mapsto (u^a)'_{b'} = (1-p t)^2u_b^a \;\; ,
a,b=1,2, \ b\not=a \ee \be\label{3-30} W_k \mapsto (W_k)'=
(1-pt)^2W_k,\, k=1,2,3. \ee Now it is easily seen that  at most five
independent absolute first-order differential invariants of the
algebra $\mathfrak{ea}_2$ can be constructed. The form of these
invariants can be taken in different  ways, however the resulting
sets of invariants will be equivalent. In the particular case, one
can suggest the form (\ref{3-28}) assuming $u^1_2-u^2_1\not=0$.

 If one assumes that  $u^1_2-u^2_1=0$ then
 the system of PDEs' in question must contain  this equation (otherwise the system
 of four PDEs will be obtained).
 To construct the other two equations,  we again use the formul{\ae}
 (\ref{3-29}--\ref{3-30}). Since  $u^1_2-u^2_1=0$, one may assume
 that $u^1_1-u^2_2 \not=0$, hence four invariants
\be\label{3-28a} \frac{W_k}{u^1_1-u^2_2}, \, k=1,2,3, \quad
\frac{u^1_2}{u^1_1-u^2_2} \ee are obtained. Finally, we arrive at
the system of the form \be\label{3-25a} g^1=0, \quad g^2=0, \quad
u^1_2-u^2_1=0, \ee where $g^1$ and $g^2$ are arbitrary smooth
functions of invariants (\ref{3-28a}). Now one easily notes  that
system (\ref{3-25a}) can formally be derived from (\ref{3-25}) as a
particular case.

Assuming $u^1_1-u^2_2 =0$, we again obtain a particular case of
the system (\ref{3-25}). This completes the proof. \hfill $\blacksquare$

Now we want to  find systems belonging to the class of systems
(\ref{3-12}), which are invariant under  the Lie algebra
$\mathfrak{ea}_3$. Such systems are described by the following
statement.

\bt A PDEs' system  of the form (\ref{3-12}) is invariant under the
Lie algebra $\mathfrak{ea}_3$
if and only if it can be reduced to the form \be\label{3-22} {\cal
B}^1=0, \quad {\cal B}^2=0, \quad {\cal B}^3=0 \ee  where ${\cal
B}^k, \, k=1,2,3$ are arbitrary smooth functions of the variables \[
W^*_{12}:=\frac{W^2_1+W^2_2}{(u^1_2 - u^2_1)^2}, \,
\,W^*_{3}:=\frac{W_3}{u^1_2 - u^2_1},
W^*:=\frac{u^1_1W^2_1+u^2_2W^2_2+(u^1_2 + u^2_1)W_1W_2}{(u^1_2 -
u^2_1)^3},\] \be\label{3-24} \frac{u^1_1 + u^2_2} {u^1_2 - u^2_1},
\quad
 \frac{U}{(u^1_2 - u^2_1)^2}
\ee
with the abbreviation $U= (u^1_1)^2 + (u^1_2)^2 + (u^2_1)^2 + (u^2_2)^2 $.
\et

\noindent \textbf{Proof.} This is  shown in a way very   similar with
respect to the previous theorem and therefore omitted here. In fact,
 the direct application of the transformations \be\label{3-6*} t
\mapsto t' = t, \;\; r_1 \mapsto r_1' =  r_1\cos p +r_2\sin p, \;\;
r_1 \mapsto r_2' =  -r_1\sin p +r_2\cos p \ee \be\label{3-6**} r_1
\mapsto (u^1)' =  u^1\cos p +u^2\sin p, \;\; r_1 \mapsto (u^2)' =
-u^1\sin p +u^2\cos p, \;\; u^3 \mapsto (u^3)' = u^3,  \ee
 generated by the operator
(\ref{3-6}) to the system of equations  (\ref{3-12}) leads to the
$\mathfrak{ea}_3$-invariant systems of the form
(\ref{3-22}--\ref{3-24}). \hfill $\blacksquare$ \noindent

Finally, using theorems 5 and 6
we can prove the theorem giving a
complete  description of PDEs systems of the form (\ref{3-8}), which
are invariant under  {\sc ecga} with the basic operators
(\ref{3-1}--\ref{3-6}).

\bt A PDEs' system  of the form (\ref{3-8}) is invariant under the
Lie algebra {\sc ecga} if and only if it possesses  the form
\be\label{3-32} {\cal H}^1=0, \quad {\cal H}^2=0, \quad  {\cal
H}^3=0, \ee where ${\cal H}^k$ are arbitrary smooth functions of the
four variables $W^*_{12}
,$
$W^*_{3}
$  and
\[ U^*:= \frac{(u^1_1 - u^2_2)^2+2(u^1_2)^2 + 2(u^2_1)^2}{(u^1_2 -
u^2_1)^2}, \quad V^*:= 2W^* - \frac{u^1_1 + u^2_2} {u^1_2 -
u^2_1}W^*_{12}
\]
\et \noindent \textbf{Proof} is based on theorem 6 and formulae
(\ref{3-20}--\ref{3-30}). We  need to find necessary and sufficient
conditions when the given system of the form
(\ref{3-22}--\ref{3-24}) admits  transformations
(\ref{3-20}--\ref{3-21}) generated by the projective operator.

 All  equations in (\ref{3-22}) have the same structure so that we can
consider  them together. Since the functions ${\cal B}^1, {\cal
B}^2$ and ${\cal B}^3$ may depend on five  variables at maximum, one
needs to find how these arguments  are transformed by the projective
transformations (\ref{3-20}--\ref{3-21}). Using formulae
(\ref{3-29}--\ref{3-30}) one easily establishes that \be\label{3-33}
\bigl(W^*_{12}\bigr)^{'}=\Bigl(\frac{W^2_1+W^2_2}{(u^1_2 -
u^2_1)^2}\Bigr)^{'}=\frac{W^2_1+W^2_2}{(u^1_2 - u^2_1)^2}=W^*_{12},
\qquad \bigl(W^*_{3}\bigr)^{'}=\Bigl(\frac{W_3}{u^1_2 -
u^2_1}\Bigr)^{'}= \frac{W_3}{u^1_2 - u^2_1}=W^*_{3}, \ee so that
$W^*_{12}$ and  $W^*_{3}$ are absolute first-order differential
invariants of {\sc ecga}. Three other   variables are transformed as
follows
 \be\label{3-34} \bigl(W^*\bigr)^{'}= W^*  -\frac{2p W^*_{12}}
{(1-pt)(u^1_2 - u^2_1)}, \ee
 \be\label{3-34a} \Bigl(\frac{u^1_1 + u^2_2} {u^1_2 -
u^2_1}\Bigr)^{'}= \frac{u^1_1 + u^2_2} {u^1_2 - u^2_1} -\frac{4p}
{(1-pt)(u^1_2 - u^2_1)}, \ee
 \be\label{3-35} \Bigl(\frac{U}{(u^1_2 -
u^2_1)^2}\Bigr)^{'}= \frac{U}{(u^1_2 - u^2_1)^2} + \frac{8p^2}
{(1-pt)^2(u^1_2 - u^2_1)^2} - \frac{4p(u^1_1 + u^2_2)} {(1-pt)(u^1_2
- u^2_1)^2}. \ee
 One observes that there is the possibility to
construct the third and fourth absolute first-order differential
invariants of {\sc ecga} using formulae (\ref{3-34}--\ref{3-35}):
\be\label{3-36} U^*:= 2\frac{U}{(u^1_2 - u^2_1)^2}-
\Bigl(\frac{u^1_1 + u^2_2} {u^1_2 - u^2_1}\Bigr)^{2} = \frac{(u^1_1
- u^2_2)^2+2(u^1_2)^2 + 2(u^2_1)^2}{(u^1_2 - u^2_1)^2} \ee and
\be\label{3-37} V^*:= 2W^* - \frac{u^1_1 + u^2_2} {u^1_2 -
u^2_1}W^*_{12}= \frac{(u^1_1 - u^2_2)(W^2_1 - W^2_2)+2(u^1_2 +
u^2_1)W_1W_2}{(u^1_2 - u^2_1)^3}.\ee

Thus, to be  invariant under transformations
(\ref{3-20}--\ref{3-21}) the equations (\ref{3-22}) must contain the
functions ${\cal B}^k = {\cal H}^k(W^*_{12}, W^*_{3}, U^*, V^*)$,
where ${\cal H}^k, \, k=1,2,3$ are  smooth functions. In the case of
arbitrary   functions ${\cal H}^k, \, k=1,2,3$, we obtain the most
general form of the first-order PDEs' system that admits  {\sc
ecga}.

This completes the proof. \hfill $\blacksquare$

\noindent \textbf{Remark 6.} {\sc ecga} can be treated as a highly
non-trivial extension of the `massless' Schr\"odinger algebra
$\mathfrak{sch}^{(0)}(2)$
by the 'exotic' operators (\ref{3-3}), and the 'mass' operator
$\Theta$. Note the last operator is produced only by the 'exotic'
operators because of the commutation relation (\ref{1-6}).

\section{Examples of non-linear systems with invariances related to the {\sc ecga}}

We now illustrate the content of the general theorems presented in
section 4 through a few examples.

\noindent \textbf{Example 1.} One of the simplest systems belonging
to the class systems, which are invariant under the 10-dimensional
Lie algebra $\mathfrak{ea}_3$, is read off from theorem~6
\be\label{4-1} W_1 = 0, \quad W_2 = 0, \quad u^1_1 + u^2_2 =0. \ee
This system can formally  be derived from (\ref{3-22}--\ref{3-24})
if one sets ${\cal B}^a= W^*_{12}, a=1,2$ and ${\cal B}^3=
\frac{u^1_1 + u^2_2} {u^1_2 - u^2_1}$. Note that the system
(\ref{4-1}) is not invariant under the projective transformations
because the last equation is incompatible with theorem 5.

The first and second PDEs in (\ref{4-1}) can be simplified using the
last equations from this system. The change of variables
$r_1\mapsto \frac{3}{2}x, r_2\mapsto \frac{3}{2} y$ and $u^3 \mapsto
\frac{3}{2} q w$ with $q\in\mathbb{R}$ brings this system to the form
\be\label{4-2}
\begin{array}{ccc} u^1_t & + u^1u^1_x + u^2u^1_y - q w_y =0 &\\
u^2_t & +  u^1u^2_x + u^2u^2_y + q w_x =0 &\\
& u^1_x  + u^2_y = 0.& \end{array} \ee It may be more appealing to
restate this in a vector notation \BEQ \label{4-3} \vec{\nabla}
\cdot \vec{v}=0 \;\; , \;\; \bigl( \partial_t
+\vec{v}\cdot\vec{\nabla}\bigr) \vec{v} - q \vec{\nabla}\wedge
\vec{\omega} = \vec{0} \EEQ where $\vec{v}=(u^1,u^2,0)$ describes
the velocity of a two-dimensional incompressible flow, $\wedge$
means vector product, $\vec{\omega}=(0,0,w)$ and
$\vec{\nabla}=(\partial_x,\partial_y,0)$.

\noindent \textbf{Remark
7.}
Equations (\ref{4-3}) can be formally obtained from the
Navier-Stokes equations, generalised to include rotational forces
\cite[eq. (2.39)]{Kreuz81}, when restricting them to a planar motion
of an incompressible fluid of density $\rho$. Because of theorem 6,
we have in addition $\vec{\nabla}\wedge \vec{v}=\vec{0}$ and if one
identifies $q=2\eta_{\rm rot}/\rho$, where $\eta_{\rm rot}$ is the
rotational viscosity, system (\ref{4-3}) is recovered.

We point out that MAI of (\ref{4-2}) is infinite-dimensional because
the system does not explicitly contain $w$  and $ w_{t}$. For
example, the system admits the operator $X_\infty
=\phi(t)\partial_{w}$ with the arbitrary given smooth function
$\phi(t)$. In fact, this operator generate the transformations
\be\label{4-4} t \mapsto t' = t, \, x \mapsto x' = x,\, y\mapsto y'
= y,
 \;\; u^a \mapsto (u^a)' =
u^a,  \;  a=1,2, \, u^3 \mapsto (u^3)' = u^3 + p\phi(t),
\ee
which preserve the form of system (\ref{4-2}).

In order to appreciate better this example, we recall briefly the
well-known shallow-water equations, of the form \cite{Pedlosky,Gill}
\be\label{4-5}
\begin{array}{ccc} u^1_t & + & u^1u^1_x + u^2u^1_y + qw_x =0\\
u^2_t & + & u^1u^2_x + u^2u^2_y +qw_y =0 \\
w_t & + & (u^1w)_x + (u^2w)_y = 0. \end{array}
\ee
The equivalent vector form is
\BEQ  \label{4-6}
\partial_t w + \vec{\nabla} \cdot \bigl( w \vec{v} \bigr) = 0 \;\; , \;\;
\bigl( \partial_t +\vec{v}\cdot\vec{\nabla}\bigr) \vec{v} + q \vec{\nabla} w = \vec{0}
\EEQ
with the same notations as above and where $\vec{v}$ is the fluid velocity, $w$ is the free surface
height over the flat bottom and the constant $q$ describes the
effect of gravity.
It should be noted that the model for the two-dimensional polytropic
gas dynamics has also form (\ref{4-5}) (see \cite{ovs} and
the references cited therein). The systems  (\ref{4-2}) and
(\ref{4-5}) have a very similar structure and trivially coincide if $w=\mbox{\rm const}$.

However, the MAI of (\ref{4-5})  is nine-dimensional with the basic
generators \cite{ovs} \be\label{3-1a} X_{-1} =-\partial_t \;\; ,
\;\; Y_{-1}^{(1)}=-\partial_{x} \;\; , Y_{-1}^{(2)} =-\partial_{y}
\;\; \ee \be\label{3-2a} Y_0^{(1)} =-t\partial_{x}-
\partial_{u^1} , \;\; Y_0^{(2)} =-t\partial_{y}- \partial_{u^2}
\ee \be\label{3-3a} X_0 = -t\partial_t - x\partial_{x}-
y\partial_{y} +u^1\partial_{u^1}+ u^2\partial_{u^2} +2 w\partial_{w}
\ee \be\label{3-5a} X_1 = -t (t\partial_t + x\partial_{x}+
x\partial_{y}) -(x-t u^1)\partial_{u^1} -(y-t u^2)\partial_{u^2} +2
t w\partial_{w} \ee \be\label{3-6a} R_0^{(12)} = -x\partial_y +
y\partial_x -  u^1\partial_{u^2} + u^2\partial_{u^1}, \ee
\be\label{3-4a} D = t\partial_t +x\partial_x +y\partial_y. \ee
This Lie algebra, which
we denote by $\wit{\mathfrak{sch}}^{(0)}(2)$, is the semi-direct sum
of the massless Schr\"odinger algebra $\mathfrak{sch}^{(0)}(2)$
and a further dilatation generator $D$, which belongs to the Cartan
subalgebra of the conformal algebra in four dimensions into which
$\mathfrak{sch}^{(0)}(2)$ is imbedded \cite{Henkel03}.

Thus, the systems (\ref{4-2}) and (\ref{4-5}) have essentially
different symmetry properties, although at first sight they appear to have a similar
structure. This structural difference implies that these two systems should correspond to different physical situations.

\medskip

\noindent \textbf{Example 2.} The simplest examples of non-linear PDEs possessing
{\sc ecga}-invariance may be read off from theorem~7. We now give two of them.
The first one reads
\be\label{4-7}
W_1 = 0, \quad W_2 = 0, \quad u^1_2 - u^2_1 =0
\ee
which in the same vector notation as in Example~1 can be written as
\BEQ
\vec{\nabla}\wedge \vec{v}=0 \;\; , \;\;
\vec{v}_t + \bigl(\vec{v}\cdot\vec{\nabla}\bigr)\vec{v} +\demi \bigl(
\vec{v}\wedge\vec{\nabla}\bigr)\wedge\vec{v}= q\, \vec{\nabla}\wedge \vec{\omega} .
\EEQ
The MAI of (\ref{4-7}) is also infinite-dimensional
because this system is invariant under the transformations (\ref{4-4}).

The second example of an {\sc ecga}-invariant system simply is
\be\label{4-8}
W_1 = 0, \quad W_2 = 0, \quad  W_3 = 0.
\ee
but apparently cannot be rendered in a simple vectorial form.
The MAI of the system (\ref{4-8}) is finite-dimensional, in
contrast to the systems listed above.

\section{Conclusions}

In this paper, the Lie and conditional symmetry methods were applied
to find non-linear PDEs  admitting the conformal Galilei algebra
{\sc cga}. Theorems 1 and 2 state that a single PDE of either first
or second order can possess this algebra only in the sense of a
conditional symmetry. However,  we have constructed a wide class of
systems of PDEs, which are invariant under the {\sc cga} and theorem
3 gives the structure of such systems.

The main part of work is devoted to the, so-called `exotic'
conformal Galilei algebra, abbreviated here by {\sc ecga}. We remind
the reader that the explicit realisation of this algebra in terms of
the first-order linear operators was found very recently
\cite{mortelly-09} so that we restricted ourselves to this. To the
best of our knowledge, there are not yet any papers devoted to
mathematically rigorous deductions of PDEs with {\sc ecga}-symmetry.
By studying the invariance of systems of second-order PDEs under
several subalgebras of {\sc ecga} (see theorems 4-7), the r\^ole of
the several possible extensions of the massless Galilei algebra
which is the common subalgebra, can be appreciated. We believe that
the most significant of our results is presented in theorem 7. If
fact, we have constructed  the most general form of  the system of
the first-order PDEs that  admits the exotic conformal Galilei
algebra.

Finally, a few examples of systems of PDEs' invariant under {\sc ecga} were presented, which
illustrate the theorems obtained, and the similarities and differences with respect to the well-known
shallow-water equations was discussed. The form of new invariant systems suggests that they might
be of interest in physical applications, for instance in magnetohydrodynamics.

\medskip

\centerline {\bf Acknowledgement}

\medskip

We thank S. Rouhani for pointing out ref. \cite{hp78}.
R.Ch. thanks the D\'epartement de Physique de la Mati\`ere et des
Mat\'eriaux, Institut Jean Lamour at the Universit\'e Henri Poincar\'e Nancy I,
where the main part of this work was carried out, for hospitality.


\end{document}